\documentclass[12pt]{article}
\usepackage{ametsoc}
\usepackage{hyperref}

\graphicspath{{./fig/}}

\newcommand{\myabstract}{
 
Extratropical extreme precipitation events are usually associated with large-scale flow disturbances, strong ascent and large latent heat release. The causal relationships between these factors are often not obvious, however, and the roles of different physical processes in producing the extreme precipitation event can be difficult to disentangle. Here, we examine the large-scale forcings and convective heating feedback in the precipitation events which caused the 2010 Pakistan flood within the Column Quasi-Geostrophic framework. A cloud-revolving model (CRM) is forced with the large-scale forcings (other than large-scale vertical motion) computed from the quasi-geostrophic omega equation with input data from a reanalysis data set, and the large-scale vertical motion is diagnosed interactively with the simulated convection. 

Numerical results show that the positive feedback of convective heating to large-scale dynamics is essential in amplifying the precipitation intensity to the observed values. Orographic lifting is the most important dynamic forcing in both events, while differential potential vorticity advection also contributes to the triggering of the first event. Horizontal moisture advection modulates the extreme events mainly by setting the environmental humidity, which modulates the amplitude of the convection's response to the dynamic forcings. When the CRM is replaced by either a single-column model (SCM) with parameterized convection or a dry model with a reduced effective static stability, the model results show substantial discrepancies compared with reanalysis data. The reasons for these discrepancies are examined, and the implications for global models and theoretical models are discussed.
}

\begin{document}
%
%

\title{\textbf{\large{Forcings and Feedbacks on Convection in the 2010 Pakistan Flood: Modeling Extreme Precipitation with Interactive Large-Scale Ascent}}}

\author{\textsc{Ji Nie}
				\thanks{\textit{Corresponding author address:} 
				 Ji Nie, Lamont-Doherty Earth Observatory, 301E Oceanography, 61 Route 9W, Palisades, NY, 10964.
				\newline{E-mail: jn2460@columbia.edu}} \\
\textit{\footnotesize{Lamont-Doherty Earth Observatory, Columbia University, New York, New York}} \\
\textsc{Daniel A. Shaevitz} \\
\textit{\footnotesize{Department of Applied Physics and Applied Mathematics, Columbia University, New York, New York}} \\
\textsc{Adam H. Sobel} \\
\textit{\footnotesize{Department of Applied Physics and Applied Mathematics, and}} \\
\textit{\footnotesize{Lamont-Doherty Earth Observatory, Columbia University, New York, New York}}
}
%
\ifthenelse{\boolean{dc}}
{
\twocolumn[
\begin{@twocolumnfalse}
\amstitle

\begin{center}
\begin{minipage}{13.0cm}
\begin{abstract}
	\myabstract
	\newline
	\begin{center}
		\rule{38mm}{0.2mm}
	\end{center}
\end{abstract}
\end{minipage}
\end{center}
\end{@twocolumnfalse}
]
}
{
\amstitle
\begin{abstract}
\myabstract
\end{abstract}
\newpage
}
  
\section{Introduction}

In late-July-early-August 2010, an unprecedented flood struck Pakistan. This devastating event, affecting about 20 million people and submerging one-fifth of Pakistan's land area under water, resulted from a series of extremely heavy rainfall events (e.g. Houze et al. 2011; also see Fig. 1a).    

The extreme precipitation events which caused the 2010 flood, as well as other similar precipitation events in the same region, have been studied by a number of investigators (e.g. Houze et al. 2011; Hong et al. 2011; Galarneau et al. 2012; Lau and Kim 2012; Martius et al. 2013; Kumar et al. 2014; Rasmussen et al. 2014; Shaevitz et al. 2016). These events usually are associated with several synoptic-scale features in the atmosphere: 1) southward intrusion of upper-level potential vorticity (PV) perturbations in the trough east of the European block causing large-scale ascent; 2) a lower-level jet associated with a monsoon depression impinging on the rising terrain in Northern Pakistan, forcing orographic ascent; 3) the same lower-level jet transporting moisture from the tropical ocean --- either the Arabian Sea, the Bay of Bengal or both --- to the flood region. From the perspective of an air column over the flood region, the first two factors can be viewed as dynamically forced large-scale ascent due to forcing at upper or lower levels, respectively. The third factor, the moisture transport, is also essential for the extreme precipitation events, in that it leads to a moist environment that favors the embedded convection. These factors, however, usually act simultaneously and interact with one another, so that the causal mechanisms are difficult to disentangle. For example, the upper-level PV anomalies can steer the lower-level jet toward the rising terrain, and the low-level southerlies cause orographic lift while simultaneously transporting moisture from the tropical oceans. The simultaneity and inter-relatedness of these varied influences obscure our understanding of their individual roles.

In these extreme precipitation events, large-scale dynamics and convection are strongly coupled. The dynamical forcings by both the upper-level PV perturbations and orographic lifting induce large-scale ascent, and stimulate deep convection by destabilizing the atmospheric stratification. The diabatic heating associated with the convection,  in turn, drives further large-scale ascent by allowing air parcels to rise across surfaces of potential temperature.  Shaevitz et al. (2016) use the quasi-geostrophic (QG) omega equation to decompose the large-scale pressure velocity (omega, $\omega$) based on reanalysis data. They show that diabatic heating accounted for more than half of the total $\omega$ in the free troposphere in the 2010 events, and that of the dynamic forcings, orographic lifting is more important than ascent due to differential advection of potential vorticity. We may think of these dynamic forcings as forcing the convection by inducing some component of the large-scale ascent, but the feedback from the convective heating onto the large-scale vertical motion must also be taken into account in order to understand these events fully. This feedback is the object of the present study.

In this study, we use the Column Quasi-Geostrophic (CQG) modeling framework, proposed by Nie and Sobel (2016), to explore the coupling between large-scale dynamics and convection. This modeling framework extends the notion of ``parameterization of large-scale dynamics" previously applied in the tropics --- including the weak-temperature-gradient method (e.g. Sobel 2000; Raymond and Zeng 2005; Daleu et al. 2015), the damped-gravity-wave method (e.g. Kuang 2008; Blossey et al. 2009; Romps 2012; Daleu et al. 2015), and related others (e.g. Mapes 2004, Bergman and Sardeshmukh 2004) --- to parameterize large-scale vertical motion. These methods allow attribution of precipitation to environmental factors such as the sea surface temperature (SST) or the large-scale tropospheric temperature profile, and have been used to aid understanding of a variety of phenomena in the tropics (e.g., Chiang and Sobel 2002; Wang et al. 2013; Wang et al. 2015; Nie and Sobel 2015; Sessions et al. 2015). In the extratropics, quasi-balanced dry adiabatic PV dynamics, captured in an approximate way by the QG system (e.g., Hoskins et al. 1985) plays a larger role in inducing large-scale vertical motion than it does in the tropics. The CQG approach parameterizes large-scale vertical motion, including components due to both the dry adiabatic QG dynamics and diabatic heating, using the QG omega equation (hereafter just ``omega equation"). The component due to diabatic heating can be thought of as representing the geostrophic adjustment processes to that heating. Combining the omega equation with a numerical model that resolves convection, the CQG approach allows one to separate the forcing and diabatic heating feedback on large-scale vertical motion, and attribute extratropical precipitation events to specific large-scale forcings.

The goal of this paper is to quantitatively untangle the interaction between the large-scale dynamics and moist convection in the 2010 Pakistan events using the CQG framework. In the first part (section 3), a cloud-resolving model (CRM) is used to simulate convection with relatively high fidelity. The large-scale forcings are obtained from reanalysis data and are used to force the CRM to produce precipitation variations in time, with the large-scale ascent modeled interactively under CQG as a function of the CRM-simulated diabatic heating as well as the imposed forcings. After showing that the precipitation of the 2010 Pakistan events is well reproduced by the CRM, we examine the positive diabatic heating feedback that makes the precipitation so intense, and quantify the role of each of the large-scale forcings in driving the events. We also explore the control of background relative humidity on the precipitation intensity. In the second part of the paper (section 4), with the CRM results serving as a benchmark, we simulate these events using two simplified, but also widely used, representations of convection. They are a single-column model (SCM) with parameterized convection, and a dry model with a reduced effective static stability, which is often used in theoretical studies to extend theories from a dry atmosphere to a moist atmosphere. We show that both of these simplified representations of convection have substantial deficiencies in reproducing the 2010 extreme events, examine the reasons causing the deficiencies, and discuss their implications. Conclusions and discussion are presented in section 5.     

\section{Methodology} 

\subsection{CQG modeling framework} 

A brief introduction to the CQG framework is presented here; more details can be found in Nie and Sobel (2016). The horizontally averaged temperature ($T$) and moisture ($q$) equations for an air column in pressure coordinates may be written as   
\begin{eqnarray}
\partial_t T  =   Adv_T + \frac{\sigma p}{R} \omega + Q,   \label{tbudget} \\
\partial_t q  =    Adv_q - s_q \omega + Q_q,  \label{qbudget}
\end{eqnarray}
where $\sigma = -\frac{R {T}}{p} \partial_p ln { \theta}$ is the dry static stability, $\omega$ is the pressure vertical velocity, and $s_q=\partial_p {q}$. $Adv_T$ and $Adv_q$ are the large-scale horizontal advection of $T$ and $q$, respectively. $Q$ and $Q_q$ are the heating and moistening tendencies due to sub-column scale processes (e.g., dry and moist convection, and radiation), respectively. 

In CQG we parameterize the large-scale dynamics using the QG$\omega$ equation (e.g. Holton 2004), assuming the horizontal structure of the disturbance
of interest has a single horizontal wavenumber ($k$, with an equivalent wavelength $L=2\pi/k$). It may be written as  
\begin{equation}
       \partial_{pp} \omega   -\sigma ( \frac{  k}{f_0})^2  \omega =  - \frac{1}{f_0}\partial_p Adv_\zeta +  \frac{R}{p}(\frac{k}{f_0})^2 Adv_T  + \frac{R}{p}  (\frac{k}{f_0})^2 Q .
       \label{eq:omega}
\end{equation}
Here $\zeta=\frac{1}{f_0}\nabla^2 \phi + f$ is the geostrophic absolute vorticity and $f_0$ is the reference value of the Coriolis parameter. The right hand side (RHS) terms are the $\omega$ forcing associated with the horizontal advection of geostrophic absolute vorticity, horizontal advection of temperature , and diabatic heating, respectively. The first two RHS terms can be combined to yield the total forcing on $\omega$ by PV advection (Trenberth 1978).  

CQG is a framework for modeling of the state of the convecting air column by, for example, a CRM. It uses the omega equation to parameterize large-scale dynamics, producing a large-scale $\omega$ field that is both a function of the convection itself and a forcing on that convection. After each CRM time step, we take the horizontally averaged diabatic heating ($Q$) computed by the CRM, and use this together with the PV forcing and orographic lift obtained from reanalysis data to calculate $\omega$ from the omega equation. The vertical temperature and moisture advection terms determined from the resulting $\omega$ and the CRM's horizontal mean $T$ and $q$ profiles are then applied in the CRM via the domain-averaged equations (Eq. 1-2). Then, the CRM proceeds to the next time step and repeats the above processes. The procedure is described visually in Fig. 1 of Nie and Sobel (2016).

The coupling between convection and large-scale dynamics through the QG$\omega$ equation provides a feedback that can amplify convective anomalies triggered by dynamics. Consider an upward $\omega$ anomaly forced by synoptic-scale influences (e.g., a PV intrusion), destabilizing the temperature stratification and leading to a local convection anomaly. Without coupling with large-scale dynamics, the convective heating anomaly will  warm the local column, drive its temperature profile towards convective stability, and shut down further convection anomalies on a convective adjustment timescale ($\sim$ 1 day). When coupled to large-scale QG dynamics, Nie and Sobel (2016) show that the convective heating anomaly induces a large-scale ascent anomaly, representing the geostrophic adjustment that tries to re-establish the geostrophic and hydrostatic balance of the large-scale flow. This ascent is associated with adiabatic cooling that compensates for a large fraction of the heating, reducing the warming and stabilization of the column, and also with moisture convergence, both of which amplify and sustain the convection anomalies and increase the adjustment time to several days. The strength of the positive feedback depends on the horizontal wavelength of the large-scale disturbance, with a maximum at wavelengths on the order of the Rossby deformation radius.
 
Since the omega equation is linear, we can express $\omega$ as the sum of three components:  
\begin{equation}
      \omega=\omega_{PV}+\omega_{BF}+\omega_Q,\label{eq:omega_decomp}
\end{equation}
where the $\omega_{PV}$, $\omega_{BF}$, and $\omega_{Q}$ are the $\omega$ profiles which would occur if only a single term were present on the RHS of (Eq. 3): PV forcing, orographic lift, and diabatic heating, respectively. The diabatic heating $Q$ is an internal, prognostic variable in CQG, so $\omega _Q$ is predicted while the other two components are imposed. Even when the diabatic heating term is the largest on the RHS of Eq. 3 as during the strong precipitation events, however, we expect time variations in vertical motion and precipitation to be controlled externally by time variations in the large-scale forcing terms. In this paper, we consider three types of large-scale forcings: PV forcing (the $Adv_\zeta$ and $Adv_T$ terms in Eq. 3), orographic lift ($\omega_0$), and horizontal moisture advection ($Adv_q$). The advection terms ($Adv_\zeta$, $Adv_T$, and $Adv_q$) are applied directly in Eq. (1-3). The orographic lifting ($\omega_0$), is not explicitly expressed in the equations, but serves as the lower boundary condition on the omega equation, applied at the top of a nominal planetary boundary layer (PBL) as discussed by Shaevitz et al. (2016). 

A significant limitation of the CQG approach (in addition to the approximations associated with the QG system itself) is that the large-scale dynamic forcings are treated as external and specified. In reality, convection can alter the large-scale flow, and thus feed back on the advection terms. We provisionally accept this limitation as a sacrifice that must be made in order to reduce a three-dimensional problem to a one-dimensional one. We view the comparison to observations made in this study, in part, as a test of the usefulness of the resulting CQG system.

\subsection{Reanalysis data } 

We use the European Centre for Medium-Range Weather Forecasts' reanalysis dataset (ERA-Interim; Dee et al. 2011) to construct the large-scale forcings and to verify the model results. The ERA reanalysis data has a temporal resolution of 6 hours and a spatial resolution of $0.7^\circ$. The precipitation data is the 12-hourly atmospheric model forecast from the same data set. The period of study is from July 1, 2010 (marked as day 0) to Aug. 10, 2010. The latitude-longitude box, $70^\circ E \sim 77^\circ E$ and $30^\circ N \sim 37^\circ N$, in which most of the rainfall fell during the 2010 events, is defined as the target region from which data are extracted for deriving the forcings and comparison to the model results. Unless otherwise specified, all results from the reanalysis are averaged over this box. 

\subsection{Cloud-resolving model} 

This subsection introduces the CRM, the System for Atmospheric Modeling (Khairoutdinov and Randall 2003), used in section 3. The other two representations of convection, the SCM and effective static stability, are introduced in section 4.

The CRM has a spatial domain of $128\ km \times 128 \ km$, 2 $km$ horizontal grid spacing, and doubly periodic horizontal boundaries. There are 64 vertical levels with vertical grid spacings stretching from 75 $m$ near the surface to about 500 $m$ in the free troposphere. The surface fluxes are computed using Monin-Obukhov similarity theory. We do not consider the effects of interactive radiation, instead applying constant radiative cooling of $-1.5\ K\ day^{-1}$ in the troposphere (levels with $T > 207.5 K$), and relaxation of temperature towards 200 $K$ over 5 days in the stratosphere, following Pauluis and Garner (2006).

The CRM domain is considered to represent the horizontal mean state of the atmospheric column in the Northern Pakistan regional box. The heterogeneities of surface conditions inside the regional box, such as the surface temperature distribution and topography, are not considered explicitly in the CRM simulation here. The orographic lifting, however, is included as an imposed lower boundary condition. In reality, the region of interest has complicated topography, with the surface pressure decreasing from 1000 $hPa$ to 500 $hPa$ from the south to the north. We approximate that as a flat surface in the CRM  with a surface pressure of 825 $hPa$, about the mean surface pressure of that regional box. We also simplify the CRM surface condition as an ocean surface with a fixed SST of $298.5\ K$, which is chosen to match the near surface air temperature in the CRM with that in the reanalysis. These idealizations in our experimental setting, although inducing some biases in the CRM mean state when compared with the reanalysis, greatly simplify the problem and remove some complexities that we hypothesize are secondary in the extreme weather events of interest. Our main interest is in the precipitation variations which occur over time. As will be seen below, the simulations produce a precipitiation time series which is reasonably similar, qualitatively and quantitatively, to those found in the reanalysis, even with these idealizations. We view this as an indication that the modeling approach captures the most essential dynamics in the extreme events.   
    
The QG$\omega$ equation is solved in the free troposphere. The upper boundary condition is a rigid lid at the  tropopause (125 $hPa$). The lower boundary condition is forced by the orographic lift, $\omega_0$, at the top of a nominal PBL, $p_0$, taken here to be $700\ hPa$.  This is high enough that frictional effects are small and the large-scale flow is close to geostrophic, and low enough that the PBL top approximately follows the orographic height (Fig. 9 in Shaevitz et al. 2016). In our simulations we simply use $\omega$ at 700 $hPa$ ($\omega_0$), obtained from the reanalysis, as orographic lift to force the CRM. Similar results are obtained, however, if we derive the $\omega _0$ from the dot product of the geostrophic wind at the top of the PBL with the topographic slope, as described by Shaevitz et al. (2016) and as shown below. In the PBL, $\omega$ is linearly interpolated between zero at the surface and $\omega_0$ at $p_0$. The  Coriolis parameter $f_0$ is set to be $7 \times 10^{-5} \ s^{-1}$, equivalent to that at $29^\circ N$. The characteristic wavelength $L$ is set to be 1500 $km$, which is close to the Rossby deformation radius at this latitude, assuming a characteristic value of the static stability. In section 3a, the choice of $L$ is justified by comparing model results with the reanalysis.  
   
The model temperature and moisture profiles are initialized with a sounding taken from a radiative-convective equilibrium (RCE) simulation performed over the same surface temperature, $298.5\ K$, and first run to a equilibrium state with the omega equation coupled, but without any externally imposed large-scale forcing. This equilibrium state has the same $T$ and $q$ profiles as the RCE profiles, a time mean $\omega$ of zero, and a mean precipitation rate of 4 $mm\ day^{-1}$. Then, beginning at a time denoted day 0, the model is forced with the large-scale forcings ($Adv_{\zeta}$, $Adv_T$, $Adv_q$ and $\omega_0$) taken from the reanalysis, starting from July 1, 2010 and continuing to August 10, 2010. Simulations from days 0-10 are discarded as spin-up.

\section{CRM Results}

Before presenting the model results, we first examine the main features of the 2010 events seen in the reanalysis. A more detailed exposition can be found in Shaevitz et al. (2016), and additional detailed studies of these events were carried out earlier by Martius et al. (2013) and Galarneau et al. (2012).
The black bars in Fig. 1a show the precipitation series in the ERA data. Standing out are the two peaks of rainfall with maxima close to 60 $mm\ day^{-1}$. The first peak, which we label ``event 1", extends from day 20 to day 23, and the second peak (``event 2") extends from day 27 to day 30. Associated with the precipitation peaks is strong upward motion (negative $\omega$) in the free troposphere (Fig. 1b), indicating a close balance between the diabatic cooling associated with ascent and the convective heating in the temperature equation. The precipitation and $\omega$ time series are the main targets of our numerical simulations.       

The large-scale forcings are shown in Fig. 2. During event 1, $Adv_{\zeta}$ and $Adv_T$ have significant maxima in the upper troposphere (Fig. 2a-b), which are indications of the PV intrusion associated with the European block in the upper stream (e.g. Lau and Kim 2012). The PV forcing, is much weaker during event 2, however, because the strong PV intrusion remained well to the north of the flood regions during event 2 (Martius et al. 2013; Shaevitz et al. 2016). The PV forcing thus plays only a secondary role in event 2.  

Horizontal moisture advection ($Adv_q$) is shown as a function of time and pressure in Fig. 2c. We also plot the column-integrated horizontal moisture advection ($\langle Adv_q \rangle$, where $\langle*\rangle$ means vertical integration through the column,) in the units of precipitation, $mm~day^{-1}$, in the lower panel. During precipitation events, $Adv_q$ itself is not a large term in the moisture budget, compared to condensation and vertical moisture advection; considering the vertical integral, $\langle Adv_q \rangle$ is much smaller than precipitation. Nonetheless, horizontal moisture advection is important when moist static energy rather than moisture alone is considered (since condensation vanishes in the moist static energy budget) and it can exert an important influence on the moisture field. During most of the examined period, $Adv_q$ is negative due to dry air import by the westerlies. However, just before the periods of heavy precipitation (days 13-18 for event 1, and days 23-26 for event 2), $Adv_q$ becomes less negative or even slightly positive due to the weakening of westerlies and strengthening of the southerlies. The positive $Adv_q$ anomalies moisten the column, making the column more favorable to the subsequent convection. During the heavily raining period, $Adv_q$ is again negative (although the moisture flux convergence is positive), because at that time the column is as moist as or even moister than the equatorward regions upwind. 

Fig. 2d shows the orographic lift ($\omega_{orog}=\vec{V}_{0} \cdot  \nabla h_s$, where $h_s$ is the orographic height and subscript $0$ indicates values at the PBL top) and $\omega$ at 700 $hPa$. The close match between them indicates that $\omega$ at the PBL top is mainly orographic forced. In our simulations, for a better consistence, we simply use $\omega$ at 700 $hPa$ ($\omega_0$) obtained from the reanalysis as orographic lift to force the CRM. The maxima in orographic lift during the two heavy precipitation events (Fig. 2d) suggest that at least some portion of extreme precipitation is orographically induced. 

\subsection{Diabatic heating feedback}
 
Next we present the CRM results. We first examine the control case, in which the model is forced by all the large-scale forcings together. The $T$ and $q$ profiles from this control case averaged between day 10 and day 40 (referred as the background profiles hereafter) are compared with the profiles obtained from the reanalysis in the same period in Fig. 3. There is a cold bias in the model results, especially in the upper troposphere, and a wet bias in the middle troposphere. Actually, the CRM relative humidity shows a similar vertical structure to that in the reanalysis data, except compressed in height, because the CRM surface level is 825 $hPa$. We presume that these discrepancies between the background profiles are due to deficiencies in the model physics, uncertainties in the reanalysis data, and the idealizations in the CRM settings (the simple radiation scheme, the flat surface, neglect of a background vertical motion, and so on). Our main interest is not in the time mean background profiles, but in simulating the precipitation variations, and the idealizations in experimental setting allow us to avoid additional complexities which would be involved in tuning the model to obtain better agreement with the time mean profiles. The most important test of whether these idealizations are adequate are the comparisons of the simulated and observed time series of precipitation and vertical motion.

The precipitation from the control simulation matches that from the reanalysis remarkably well (Fig. 1a). The two extreme events are well reproduced by the model, except that the model results underestimate the precipitation intensity slightly. The model results even capture the minor precipitation peaks around days 13, 18, and 34. The simulated $\omega$ also largely resembles that in the reanalysis data (Fig. 1b-c). 
 
The components of $\omega$ ($\omega_{PV}$, $\omega_{BF}$, and $\omega_{Q}$) in the reanalysis and model results are compared side by side in Fig. 4. The $\omega$ components in the reanalysis are first computed using full omega equation on the 3-dimensional sphere grids before averaging over the regional box (Shaevitz et al. 2016). Thus, this diagnosis does not assume a specific disturbance wavelength. Further, this analysis treats the diabatic heating as known (the diabatic heating in the reanalysis is calculated as a residual from the temperature  equation) and compares its influence in the omega equation to that of the other terms. In reality, these terms are not independent; the diabatic heating is related to the large-scale processes represented by the other terms, as they influence the convection. These influences are considered explicitly in the next section. Here we simply establish the magnitudes of the different terms in the omega equation.

 The sum of the $\omega$ components obtained from the 3-dimensional QG$\omega$ calculation (left panels in Fig. 4) is very close to the $\omega$ directly obtained from the reanalysis (Fig. 1b). Consistent with Fig. 2, considerable ascent in the upper troposphere is directly forced by PV forcing during event 1 (Fig. 4a). A large portion of the ascent in the lower troposphere is associated with orographic lift (Fig. 4c). The $\omega$ component due to diabatic heating (Fig. 4e) is largest in the middle troposphere, and has the largest amplitude among the three components during the events. 

The $\omega$ components in the CQG model (right panels in Fig. 4) are calculated differently from those described above (and shown in the left panels of Fig. 4). Using the regional-box-averaged large-scale forcings in the reanalysis and $Q$ in the CRM, we solve the single wavenumber QG$\omega$ equation (Eq. 3) with the specified characteristic wavenumber $k$. Thus, comparing the dynamic forcing components of $\omega$ can guide the choice of $k$. Fig. 4a-d show that $\omega_{PV}$ and $\omega_{BF}$ in the model results approximately match those in the reanalysis, indicating that $k=4.2  \times 10^{-6} m^{-1}$ (equivalent to 1500 $km$ wavelength) used here is reasonable. 

In the model, the component of $\omega$ due to heating ($\omega_Q$) is related to the large-scale forcings through their influence on convection. The large-scale forcings alter the $T$ and $q$ profiles in the column, which then determine the embedded convection and convective heating ($Q$). Meanwhile, $\omega_Q$ itself also modifies $T$ and $q$ interactively. Thus a good simulation of $\omega_Q$ requires a good simulation of convection as well as a good representation of the coupling between convection and the large-scale dynamics. The agreement between $\omega_Q$ in the reanalysis and model results (Fig. 4e-f) indicates that the CRM with the CQG framework captures the above processes and reproduces the 2010 extreme events well. 
    
Similarly, we can decompose the precipitation variations to different components of the forcing and the response. With the approximation that the adiabatic cooling due to ascent is balanced by latent heating due to condensation, we can estimate the precipitation associated with $\omega$ or its individual components, 
\begin{equation}
P \approx - \frac{c_p}{gL_c} \langle \frac{\sigma p}{R} \omega \rangle,
\label{eq:p_omega}
\end{equation}
where $c_p$ is the specific heat of air at constant pressure, and $L_c$ is the latent heat of condensation. Precipitation estimated in this way allows negative values with descending $\omega$. 
The precipitation decomposition is shown in Fig. 5. Each color line is calculated using Eq. 5 with $\omega$ replaced by one of the components shown in Fig. 4. In the reanalysis, using Eq. 5 underestimates the precipitation, as shown by the black dashed line compared to the black solid line in Fig. 5a. This may be due to errors in the approximation 
(\ref{eq:p_omega}), or possibly to the uncertainties in the reanalysis precipitation; in the ERA reanalysis, as with any data set generated with data assimilation, the temperature and moisture budgets do not close due to analysis increments. The precipitation in the TRMM data set is smaller than that in the ERA reanalysis (Fig. 5 in Martius et al. 2013). In the model, the total precipitation estimated by Eq. 5 is very close to that directly output by the model. Nevertheless, the precipitation decomposition is consistent with the $\omega$ decompositions in Fig. 4. It shows the significant roles of PV forcing in event 1 and of orographic lift in both events 1 and 2, and the large contributions from the diabatic heating feedback component. The model results again show precipitation decomposition similar to the reanalysis diagnoses.  

The diabatic heating feedback of convection to the large-scale dynamics as parameterized in CQG is essential for the CRM to reproduce the extreme events. To demonstrate this, we run an experiment in which the diabatic heating feedback is turned off, by removing the third RHS term in Eq. 3, while keeping others the same as in the control case. This modified case produces precipitation (shown by black circles in Fig. 5b) that is much smaller than that in the control case. Without the positive feedback of diabatic heating on large-scale dynamics, convection only responds passively, producing approximately just enough diabatic heating to balance the imposed large-scale forcings.

\subsection{Attribution of precipitation to large-scale forcings}

In the previous sections we treated the diabatic heating as a known quantity and compared it directly to other processes in the omega equation. Here we attribute the precipitation variations to the large-scale forcings. We consider the role of each forcing in modulating the convection, and thus diabatic heating, while treating the diabatic heating itself as an internal, prognostic quantity associated with the convection simulated by the CRM. Three types of large-scale forcing (Fig. 2) are examined: the upper-level PV forcing, the lower-level orographic lift, and horizontal moisture advection. The first two influence convection dynamically by forcing ascent that destabilizes the atmospheric stratification. The moisture advection influences convection thermodynamically by modifying the humidity of the column. Nie and Sobel (2016) showed that the responses of convection vary with the type of forcing as well as the forcing altitude. 

The attribution of precipitation to each of the large-scale forcings is examined by forcing the model with one of the forcings at a time. Note that while the QG$\omega$ equation is linear, so that the different components of $\omega$ must sum up to the total diagnosed by that equation, the precipitation or $\omega$ time series from these individual forcing experiments need not add up to that from the control experiment, because the simulated convection need not be a linear function of the forcings. Precipitation from these experiments is shown as the red lines in Fig. 6b-d. For reference, also plotted are the control case precipitation (black line) and the precipitation component corresponding to the imposed large-scale forcing (blue line). With only PV forcing (Fig. 6b), the model results capture about half of the control case precipitation in event 1. Event 2 is missed since PV forcing is very small there. When forced only by orographic lift (Fig. 6c), the model captures the other half of precipitation in event 1 and almost all of the precipitation in event 2. The low-level orographic lift seems to be more important in causing heavy precipitation than is the upper-level PV forcing, especially for event 2. A possible reason may be that convection is more sensitive to lower tropospheric temperature and moisture perturbations than to upper tropospheric ones (e.g. Kuang 2010; Tulich and Mapes 2010; Nie and Kuang 2012). 

Horizontal moisture advection affects convection in these events in a different way. When the model is forced only by $Adv_q$ (Fig. 6d), the simulated precipitation does $\it{not}$ resemble that in the control case. Instead, it is closely correlated with the moisture forcing ($\langle Adv_q \rangle$, in the same units as the precipitation, shown as the blue line in Fig. 6c.) During most of the simulation, the negative $Adv_q$ dries the column, leading to dry periods with little precipitation between precipitation peaks. However, the sharp precipitation peaks at days 17, 26, and 34 closely follow the periods of positive $Adv_q$ with a time delay of about one day. This experiment shows that  even without the dynamic forcings $Adv_q$ variations on the synoptic timescale can trigger precipitation events directly. However, this direct triggering effect is not expressed in the 2010 events. Rather, as will discuss next, $Adv_q$ exerts its influence mainly by modifying the environment to a drier or wetter state, which affects the strength of the convection's response to the dynamic forcings.     

The sum of the precipitation in the individually-forced experiments is compared with that in the control case precipitation in Fig. 6a. An offset of 8 $mm\ day^{-1}$ is subtracted, to account for the multiple counting of the component of the time-mean precipitation needed to balance the radiative cooling (corresponding to 4 $mm\ day^{-1}$). Since we are adding results from three simulations, it is appropriate to subtract twice this amount, 8 $mm\ day^{-1}$, before comparing to the control case.

Even apart from this mean offset, the sum of the individually-forced experiments need not equal the 
control experiment because the system is not inherently linear.
During the two precipitation peak periods, the two are in fact approximately equal, indicating that the strong dynamical forcings account for most of the precipitation there. 
However, there are some differences between the sum of the individually-forced experiments and the control case. We hypothesize that these are due to interactions between the forcings. For example, around days 11$\sim$12, a dynamic destabilizing coincided with strong drying by horizontal advection. As a result, the onset of precipitation is delayed to day 13$\sim$14, by when the drying is much weaker. As another example, the moistening by horizontal advection at day 17 and 26 coincided with dynamically-forced descent, mostly due to the PV forcing, so that the horizontal moisture advection-forced precipitation seen in Fig. 6d is suppressed in the control case. 

\subsection{Influence of environmental relative humidity on convection}

Environmental humidity exerts an important control on convection (e.g., Raymond 2000; Derbyshire et al. 2004; Bretherton et al. 2004; Sobel et al. 2004; Holloway and Neelin 2009). In section 3b, we examined the effects of synoptic-scale variations in horizontal moisture advection on the extreme events. In this subsection, we focus on 
the control on precipitation exerted by background relative humidity on longer time scales.

We perform experiments that are same with the control case, but with the background humidity nudged towards different states. We perform four experiments, in which $q$ is nudged to the control case $q$ multiplied by a factor of 0.6 (Dry0.6), 0.8 (Dry0.8), 1.1 (Moist1.1), and 1.2 (Moist1.2), respectively. To rule out effects due to the associated temperature changes, the $T$ profiles are nudged to the control case $T$ in these runs. The nudging timescale is 5 days, so that the synoptic timescale variations are not strongly affected by the nudging. These cases are forced with the same large-scale forcings as in the control case. Fig. 3 shows the resulting background $T$ and $q$ of the driest and the moistest case. The $T$ profiles are very close to each other, while the $q$ and relative humidity profiles deviate from each other as designed. Thus, these experiments shall be viewed as examining the dependence of extreme precipitation on background relative humidity.  Because the nudging is relatively weak, the $q$ differences between these runs are much smaller than the differences in the target profiles towards which they are nudged. 

The results of these experiments show that the precipitation in the extreme events is very sensitive to the background   humidity (Fig. 7a). The shapes of the precipitation time series of these cases are similar, but the difference in the magnitudes of the precipitation maxima between the driest case and the moistest case is more than a factor of 3. The dependence of extreme precipitation intensity on background humidity is summarized by a scatter plot in Fig. 7c, with the data averaged in the same way. The $x$-axis is the background precipitable water (column-integrated specific humidity, averaged over day 10$\sim$40), and the $y$-axis is the precipitation rate averaged over the two extreme events (day 20$\sim$23 and day 27$\sim$30). Over the examined range, the precipitation is linearly proportional to the background precipitable water, with a slope of 1.68 ${d}^{-1}$ (equivalently, $mm\ d^{-1}\ mm^{-1}$). This strong dependence suggests that the intensities of extreme events can be modulated by large-scale circulation anomalies that induce regional background relative humidity changes, such as those due to the El Ni$\tilde{n}$o Southern Oscillation (e.g. Lau and Kim 2012), monsoon (e.g. Freychet et al. 2015); or anthropogenic climate change (e.g. Allan and Soden 2008; Byrne and O'Gorman 2013). 

\section{Results with simplified representations of convection}
 
\subsection{Results using a convective parameterization}

We repeat the experiments in section 3 with the CRM replaced by the MIT SCM (Emanuel and $\breve{Z}$ivkovic-Rothman 1999). The core of this SCM is Emanuel's convective parameterization (Emanuel 1991), which represents an ensemble of convective clouds, and a stratiform cloud parameterization (Bony and Emanuel 2001). The vertical resolution is 25 $hP$, and the time step is 5 minutes. All other experimental settings, such as the radiative cooling, surface conditions, and others, are the same as in the CRM experiments. 

The precipitation from the SCM  in the control case, in which the SCM is forced by all the large-scale forcings, is compared with the reanalysis precipitation in Fig. 8a. The model captures the general shape of the precipitation time series, but underestimates the intensities of the two main events by as much as half, indicating that the SCM responds too weakly to the large-scale forcings.  

As we did above for the CRM, we force the SCM with the large-scale forcings one at a time to examine their individual effects. These SCM results, shown in Fig. 8b-d, can be directly compared with the CRM results in Fig. 6b-d (Note the range of y-axis in Fig. 8b-d is stretched for better visibility.) For the orographic lift forcing, the SCM precipitation is comparable with that from the CRM (Fig. 8c, Fig. 6c). However, when forced by PV forcing, the SCM produces much weaker precipitation than does the CRM (Fig. 8b, Fig. 6b), suggesting that the SCM does not respond strongly enough to the dynamic destabilization in the upper troposphere. The SCM also shows much weaker precipitation variations in response to synoptic timescale moisture forcing than does the CRM (Fig. 8d, Fig. 6d).  

The lack of convective moisture sensitivity in SCM is also shown in its responses to background relative humidity changes. The experiments with the background $q$ profiles nudged to different states, as in section 3c with the CRM, are performed with the SCM. The SCM results again show much weaker dependence on the background $q$ than did the CRM (Fig. 7a-b). We summarize all the SCM cases on the same precipitable-water-and-peak-precipitation scatter plot with the CRM results in Fig. 7c. A linear fit of all the SCM cases examined here gives the dependence of extreme precipitation intensity on background precipitable water with a slope of 0.89 $mm/day\ per\ mm$, about half of the CRM results' slope. The linear fits to the CRM and SCM results converge in the dry limit where latent heating is negligible. However, the discrepancies between them increase as the environment becomes moister. 

Although only one convective parameterization is examined here, the lack of convective sensitivity to moisture is a common problem in many convective parameterizations (e.g. Derbyshire et al. 2004). This deficiency may affect the modeling of extreme precipitation in general circulation models (GCMs) which use such parameterizations.

\subsection{Results using an effective static stability}
 
In some theoretical studies, the effects of latent heating are taken into account in by replacing the dry static stability $\sigma$ by an effective (or moist) static stability $\sigma_{e}$ in otherwise dry models (e.g. Kiladis et al. 2009; O'Gorman 2011; Cohen and Boos 2016).
In the QG$\omega$ equation, one may define $\sigma_{e}$ as 
\begin{equation}
       \sigma_{e} =  \sigma   + \frac{R}{p \omega}  Q,
       \label{eq:sigma_e}
\end{equation} 
and rewrite Eq. 3 as
\begin{equation}
\partial_{pp} \omega   -\sigma_e ( \frac{  k}{f_0})^2  \omega =  - \frac{1}{f_0}\partial_p Adv_\zeta +  \frac{R}{p}(\frac{k}{f_0})^2 Adv_T.  
\label{eq:omega_sigma_e}
\end{equation} 
In Eq. 7, the latent heating $Q$ is hidden in the definition of $\sigma_e$, but the content of the equation is formally unchanged. The new parameter $\sigma_e$ absorbs all the complexities of convection and should, in principle, depend on the state of the column. However, in studies where this device is used, it is common to further specify that $\sigma_e$ is constant fraction of $\sigma$ (e.g. Emanuel et al. 1994; Cohen and Boos 2016):
\begin{equation}
  \sigma_{e}  =  \begin{cases}
    \sigma & : \omega \ge 0, \\
     \alpha  \sigma & : \omega<0,
  \end{cases} 
  \end{equation}
where $\alpha \in [0, 1]$ is a free parameter. $\alpha=1$ represents a dry atmosphere, and $\alpha=0$ means that the atmosphere is exactly neutral to moist convection, such as would be the case for a saturated atmosphere with a moist adiabatic temperature profile. For any value of $\alpha$ less than unity, the reduction in static stability is conditional on the sign of $\omega$, so that $\sigma_e < \sigma$ only when there is large-scale ascent. The use of $\sigma_e$ with this parameterization removes any dependence of convection on moisture.

We force Eq. 7 with the same PV forcing and orographic lift as before ($Adv_q$ is not applicable here, since there is no explicit coupling to the moisture field). The computed $\omega$ includes both the forcing component and the diabatic heating component, with the latter implicitly expressed via Eq. 7. The resulting $\omega$ is then converted into precipitation using Eq. 5. This precipitation, computed with different values of $\alpha$, is compared with the reanalysis precipitation in Fig. 9a. Precipitation in the dry limit ($\alpha=1$) corresponds solely to that produced by the dynamic forcings, with no diabatic feedback. As expected, precipitation increases as $\alpha$ decreases. The value $\alpha=0.2$ produces good results for event 1, but greatly underestimates precipitation in event 2. As $\alpha$ further decreases, precipitation in event 1 increases sharply, while precipitation in event 2 increases only slowly. 
 
By tuning $\alpha$, we can easily change the amplitude of the response to PV forcing, but not the response to orographic lift. Fig. 9b-e shows the results with Eq. 8 being forced by PV forcing and orographic lift separately. The increase of precipitation with decreasing $\alpha$ seen in Fig. 9a is almost entirely due to the increase of the PV-forced component (Fig. 9b). The precipitation response to orographic lift is $\it{not}$ sensitive to $\alpha$ (Fig. 9c). In Fig. 10d, we show the $\omega$ profiles forced by PV forcing on day 22, a representative day during event 1. With $\alpha=0.2$, the dry system (Eq. 7) matches the CRM results well. Decreasing $\alpha$ further leads to a large overestimation of the PV-forced $\omega$. Fig. 9e shows the $\omega$ profiles forced by orographic lift on day 28, a representative day during event 2. Above the PBL top, $\omega$ decreases with height. This is qualitatively different from the CRM results, which show that orographic lift can trigger convection whose latent heating maximizes in the free troposphere (black dashed line in Fig. 9e), and which agree well with the reanalysis data (Fig. 4e).  

This qualitative difference can be understood by considering the analytic solution to (\ref{eq:omega_sigma_e}) for constant $\sigma_e$, in the special case that the PV forcing (the RHS of Eq. 7) is zero, but there is a nonzero orographic lift at the lower boundary, $\omega_0$. The analytic solution, with upper boundary condition of $\omega(p_t)=0$, is 
\begin{eqnarray}
\omega=\omega_0 (1-e^{ 2 \frac{k}{f}\sqrt{\sigma_e} (p_t-p_0)} )^{-1}  (e^{ \frac{k}{f}\sqrt{\sigma_e} (p-p_0)} -e^{ \frac{k}{f}\sqrt{\sigma_e} (2p_t-p-p_0)}  ) \nonumber \\
\approx \omega_{0} e^{ \frac{k}{f}\sqrt{\sigma_e} (p-p_{0})},
\end{eqnarray}
where $p_0$ is the pressure at the top of the PBL and $p_t$ is the pressure at the tropopause. We see that the solution must decay exponentially with height. A smaller $\alpha$ leads to only slower decay. The use of $\sigma_e$ (Eq. 8) cannot produce an $\omega$ profile with a maximum in the interior as observed. 

With a sufficiently complex representation of $\sigma _e$, allowing it to be a function of pressure (and to take on negative values), the parameterization (Eq. 7) can in principle capture any solution. However, in that case this parameterization is of little value, since then one is back to the problem of parameterizing convection in its full complexity. Our conclusion is that while the dry system with an effective static stability may be useful for some purposes, it is unsatisfactory for the purpose of modeling the events studied here, and perhaps any deep convective events triggered by orographic forcing.

\section{Conclusions and Discussion}

This study examines the large-scale forcings and the convective responses to those forcings in the extreme precipitation events of 2010 in Pakistan, using the Column Quasi-Geostrophic modeling framework. Forcing a CRM with large-scale forcings taken from reanalysis data and coupling large-scale dynamics with convection through the QG$\omega$ equation, we have successfully reproduced the main features of the extreme precipitation events. The positive feedback of convective heating to large-scale vertical motion is essential in amplifying the precipitation intensity to reach the observed values. The large-scale forcings, including those due to differential vorticity and temperature advection (which together can be represented as differential advection of potential vorticity), mechanical orographic lift, and horizontal moisture advection, play significant roles forcing convection. Orographic lift is the most important dynamical forcing, especially for the second of the two heavy precipitation events, while PV forcing also contributes to the triggering of the first event. Horizontal moisture advection modulates the convection mainly by setting the environment humidity, which controls the amplitude of convection's responses to the dynamic forcings. 
  
We also performed calculations using CQG in which convection was represented by a convective parameterization in a single column model (SCM) as well as by an effective static stability in a dry model.  The results from these calculations have more substantial deficiencies than do the CRM results. The SCM underestimates the convective response to upper-level PV forcing, and shows much weaker sensitivity to environmental humidity than does the CRM.  The effective static stability can be tuned to match the extreme precipitation triggered by upper-level PV forcing, but its response to orographic is qualitatively inaccurate, in that it cannot produce maximum ascent in the interior whereas the observations and CRM simulations both show such maxima.  In addition, the effective static stability cannot represent the dependence of convection on environmental humidity. 

The method used here can be used to investigate other extreme precipitation events. Shaevitz et al. (2016) find that the extreme precipitation events of summer 2014 in India and Pakistan are similar to the second event of the 2010 flood studied here. They also find a strong correlation between the extreme precipitation intensity to orographic lift and column precipitable water in a 10-years reanalysis data in this region (their Fig. 14), suggesting that our findings may apply to a larger number of events. Globally, a large number of the most intense precipitation events occur in the subtropical belt during summer (Zipser et al. 2006; Cecil and Blankenship 2012), likely results of interactions between extratropical dynamics with sufficient moisture supply from the tropics. Applying the CQG method of this study to events in other regions, e.g., Texas (e.g. Wang et al. 2015) or the Middle East (e.g. de Vries et al. 2013), may be useful in identifying the most important physical factors leading to extreme precipitation events in these different regions. 
 
Climate simulations with GCMs have been shown to underestimate extreme precipitation, and to exhibit significant uncertainties in the changes in extreme precipitation they project as the climate warms (e.g. Sun et al. 2006, 2007; O'Gorman 2015), mostly due to deficiencies in convection parameterizations. Results of this study suggest that the lack of convective moisture sensitivity in convective parameterizations (e.g., Derbyshire et al. 2004) may contribute to the biases. The CQG approach may also be used to study the physics underlying the projected changes of extreme precipitation in the future. Variations in the large-scale vertical motion profiles between different models have been shown to be responsible for much of the diversity in GCM projections (O'Gorman and Schneider 2009; Sugiyama et al. 2010). The CQG approach allows us to study in detail how different factors influence vertical motion profiles in the presence of convection, and may provide a route to some insight on this problem.

\begin{acknowledgment} 
The authors thank Shuguang Wang, William Boos, and Zhiming Kuang for discussion. This research was supported by the Lamont Postdoctoral Fellowship to JN, an AXA Award from the AXA Research Fund to AHS, and NASA grant XXX which supported DAS.
\end{acknowledgment}


\ifthenelse{\boolean{dc}}
{}
{\clearpage}
\bibliographystyle{ametsoc}
\bibliography{references}

\newpage

\begin{center}  REFERENCE:  \end{center}

Allan, R. P. and B. J. Soden, (2008): Atmospheric warming and the amplification of precipitation extremes. Science, 321, 1481--1484.

Bergman, J. W. and P. D. Sardeshmukh, (2004): Dynamic Stabilization of Atmospheric Single Column Models. J. Clim., 17, 1004--1021.

Blossey, P. N., C. S. Bretherton, and M. C. Wyant, (2009): Understanding subtropical low cloud response to a warmer climate in a superparameterized climate model. Part II: Column modeling with a cloud-resolving model. J. Adv. Model. Earth Syst.,1, doi:10.3894/JAMES.2009.1.8.

Bony, S. and K. A. Emanuel, (2001): A parameterization of the cloudiness associated with cumulus convection: Evaluation using TOGA COARE data. J. Atmos. Sci., 58, 3158--3183.

Bretherton, C. S., M. E. Peters, and L. E. Back, (2004): Relationships between water vapor path and precipitation over the tropical oceans. J. Climate, 17, 1517--1528.

Byrne, M. P., and O'Gorman, P. A., (2013): Land-ocean warming contrast over a wide range of climates: convective quasi-equilibrium theory and idealized simulations, J. Climate, 26, 4000--4016. 

Cecil, D. J. and C. B. Blankenship, (2012): Toward a Global Climatology of Severe Hailstorms as Estimated by Satellite Passive Microwave Imagers. J. Climate, 25, 687--703.

Chiang, J. C. and A. H. Sobel, (2002): Tropical tropospheric temperature variations caused by ENSO and their influence on the remote tropical climate, J. Climate, 15, 2616--2631.

Cohen, N. Y., and W. R. Boos, (2016): Perspectives on moist baroclinic instability: implications for the growth of monsoon depressions. J. Atmos. Sci., In press. doi: http://dx.doi.org/10.1175/JAS-D-15-0254.1.

Dee, D. P., et al., (2011): The ERA-Interim reanalysis: Configuration and performance of the data assimilation system. Q. J. R. Meteorol. Soc. 137: 553--597.

Derbyshire, S. H., I. Beau, P. Bechtold, J. Y. Grandpeix, J. M. Piriou, J. L. Redelsperger, and P. M. M. Soares, (2004): Sensitivity of moist convection to environmental humidity. Quart. J. Roy. Meteor. Soc., 130, 3055--3079.

de Vries, A. J., E. Tyrlis, D. Edry, S. O. Krichak, B. Steil, and J. Lelieveld, (2013): Extreme precipitation events in the Middle East: Dynamics of the Active Red Sea Trough. J. Geophys. Res. Atmos., 118, 7087--7108, doi:10.1002/jgrd.50569.

Emanuel, K. A., (1991): A scheme for representing cumulus convection in large-scale models. J. Atmos. Sci., 48, 2313--2335.

Emanuel, K. A., J. D. Neelin, and C. S. Bretherton, (1994): On large-scale circulations in convecting atmospheres. Q. J. R. Meteorol. Soc., 120, 1111--1143.

Emanuel, K. A., and M. $\breve{Z}$ivkovic-Rothman, (1999): Development and evaluation of a convection scheme for use in climate models. J. Atmos. Sci., 56, 1766--1782.

Freychet, N., H. H. Hsu, C. Chou, and C.-H. Wu, (2015): Asian Summer Monsoon in CMIP5 Projections: A Link between the Change in Extreme Precipitation and Monsoon Dynamics. J. Climate, 28, 1477--1493.

Galarneau Jr., T. J., T. M. Hamill, R. M. Dole, and J. Perlwitz, (2012): A Multiscale Analysis of the Extreme Weather Events over Western Russia and Northern Pakistan during July 2010. Mon. Wea. Rev., 140, 1639--1664.

Holloway, C. E. and J. D. Neelin, (2009): Moisture Vertical Structure, Column Water Vapor, and Tropical Deep Convection. J. Atmos. Sci., 66, 1665--1683.

Holton, J., (2004): An introduction to dynamic meteo- rology. Forth edition, Elsevier Academic Press. P164--168.

Hong, C. C., H. H. Hsu, N. H. Lin, and H. Chiu, (2011): Roles of European blocking and tropical extratropical interaction in the 2010 Pakistan flooding, Geophys. Res. Lett., 38, L13806, doi:10.1029/2011GL047583.


Hoskins, B. J., M. E. Mcintyre, and A. W. Robertson, (1985): On the use and significance of isentropic potential vorticity maps, Q. J. R. Meteorol. Soc., 111, 877--946.

Houze Jr., R. A., K. L. Rasmussen, S. Medina, S. R. Brodzik, and U. Romatschke, (2011): Anomalous Atmospheric Events Leading to the Summer 2010 Floods in Pakistan. Bull. Amer. Meteor. Soc., 92, 291--298.

Khairoutdinov, M. F. and D. A. Randall, (2003): Cloud resolving modeling of the ARM summer 1997 IOP: Model formulation, results, uncertainties, and sensitivities. J. Atmos. Sci., 60, 607--625.

Kiladis, G. N., M. C. Wheeler, P. T. Haertel, K. H. Straub, and P. E. Roundy, 2009: Convectively coupled equatorial waves. Rev. Geophys., 47, RG2003.

Kuang, Z., (2008): Modeling the interaction between cumulus convection and linear waves using a limited domain cloud system resolving model. J. Atmos. Sci., 65, 576--591.

Kuang, Z., (2010): Linear response functions of a cumu- lus ensemble to temperature and moisture perturbations and implication to the dynamics of convectively coupled waves. J. Atmos. Sci., 67, 941--962.


Lau, W. K. M. and K. Kim, (2012): The 2010 Pakistan Flood and Russian Heat Wave: Teleconnection of Hydrometeorological Extremes. J. Hydrometeor, 13, 392--403.


Mapes, B. E., (2004): Sensitivities of cumulus ensemble rainfall in a cloud-resolving model with parameterized large-scale dynamics. J. Atmos. Sci., 61, 2308--2317.

Martius, O.,  H. Sodemann, H. Joos, S. Pfahl, A. Winschall, M. Croci-Maspoli, M. Graf, E. Madonna, B. Mueller, S. Schemm, J. Sedl$\acute{a}\breve{c}$ek, M. Sprenger, H. Wernli, (2013): The role of upper-level dynamics and surface processes for the Pakistan flood of July 2010. Q. J. R. Meteorol. Soc. 139: 1780--1797. 

Nie, J. and Z. Kuang, (2012): Responses of shallow cumulus convection to large-scale temperature and moisture perturbations: a comparison of large-eddy simulations and a convective parameterization based on stochastically en- training parcels, J. Atmos. Sci., 69, 1936--1956.

Nie, J. and A. H. Sobel, (2015): Responses of tropical deep convection to the QBO: cloud-resolving simulations, J. Atmos. Sci., 72, 3625--3638.

Nie, J. and A. H. Sobel, (2016): Modeling the Interaction between Quasi-Geostrophic Vertical Motion and Convection in a Single Column, J. Atmos. Sci., 73, 1101--1117.

O'Gorman, P. A. and T. Schneider, (2009): The physical basis for increases in precipitation extremes in simulations of 21st-century climate change. Proc. Nat. Acad. Sci., 106, 14773--14777.

O'Gorman, P. A., (2011): The Effective Static Stability Experienced by Eddies in a Moist Atmosphere. J. Atmos. Sci., 68, 75--90.

O'Gorman, P. A., (2015): Precipitation extremes under climate change, Current Climate Change Reports, 1, 49--59. 

Pauluis, O. and S. Garner, (2006): Sensitivity of radiative-convective equilibrium simu- lations to horizontal resolution, J. Atmos. Sci., 63, 1910--1923.

Rasmussen, K. L., A. J. Hill, V. E. Toma, M. D. Zuluaga, P. J. Webster, and R. A. Houze, (2014): Multiscale analysis of three consecutive years of anomalous flooding in Pakistan. Q.J.R. Meteorol. Soc., 141: 1259--1276.

Raymond, D. J., (2000): Thermodynamic control of tropical rainfall, Quart. J. Roy. Meteor. Soc., 126, 889--898.

Raymond, D. J. and X. Zeng, (2005): Modelling tropical atmospheric convection in the context of the weak temperature gradient approximation. Q.J.R. Meteorol. Soc., 131: 1301--1320.

Romps D. M., (2012): Weak Pressure Gradient Approximation and Its Analytical Solutions. J. Atmos. Sci., 69, 2835--2845.

Shaevitz, D. A., J. Nie and A. H. Sobel, (2016): The 2010 and 2014 floods in India and Pakistan: dynamical influences on the vertical motion and precipitation. Submitted to J. Geophys. Res.. The manuscript is available at \url{http://www.ldeo.columbia.edu/~sobel/Papers/shaevitz_et_al_QJ.pdf}

Sessions, S. L., M. J. Herman, and S. Senti$\acute{c}$, (2015): Convective response to changes in the thermodynamic environment in idealized weak temperature gradient simulations, J. Adv. Model. Earth Syst., 7, 712--738, doi:10.1002/2015MS000446.

Sobel, A. H. and C. S. Bretherton, (2000): Modeling tropical precipitation in a single column. J. Clim., 13, 4378--4392.

Sobel, A. H., S. E. Yuter, C. S. Bretherton, and G. N. Kiladis, (2004): Large-scale meteorology and deep convection during TRMM KWAJEX.  Mon. Wea. Rev.,132, 422--444.

Sugiyama, M., H. Shiogama, M. Emori, (2010): Precipitation extreme changes exceeding moisture content increases in MIROC and IPCC climate models. Proc. Nat. Acad. Sci., 107, 571--575, doi: 10.1073/pnas.0903186107.

Sun, Y., S. Solomon, A. Dai, and R. W. Portmann, (2006): How often does it rain? J. Climate, 19, 916--934.

Sun, Y., S. Solomon, A. Dai, and R. W. Portmann, (2007): How Often Will It Rain?. J. Climate, 20, 4801--4818.

Trenberth, K. E., (1978): Interpretation of the diagnos- tic quasi-geostrophic omega equation. Mon. Wea. Rev., 106, 131-137.

Tulich, S. N. and B. E. Mapes, (2010): Transient envi- ronmental sensitivities of explicitly simulated tropical con- vection. J. Atmos. Sci., 67, 957--974.

Wang, S., A. H. Sobel, and Z. Kuang, (2013): Cloud-resolving simulation of TOGA COARE using parameterized large-scale dynamics. J. Geophys. Res., DOI: 10.1002/jgrd.50510. 

Wang, S., A. H. Sobel, and J. Nie, (2015): Modeling the MJO rain rates using parameterized large scale dynamics: vertical structure, radiation, and horizontal advection of dry air, submitted to J. Adv. Model. Earth Syst.

Wang, S. S.-Y., W.-R. Huang, H.-H. Hsu, and R. R. Gillies, (2015): Role of the strengthened El Ni$\tilde{n}$o teleconnection in the May 2015 floods over the southern Great Plains, Geophys. Res. Lett., 42, 8140--8146, doi:10.1002/2015GL065211.


Zipser, E. J., C. Liu, D. J. Cecil, S. W. Nesbitt, and D. P. Yorty, (2006): Where are the most intense thunderstorms on earth? Bull. Amer. Meteor. Soc., 87, 1057--1071.


\begin{figure}[htbp] 
   \centering
   \includegraphics[width=1 \textwidth]{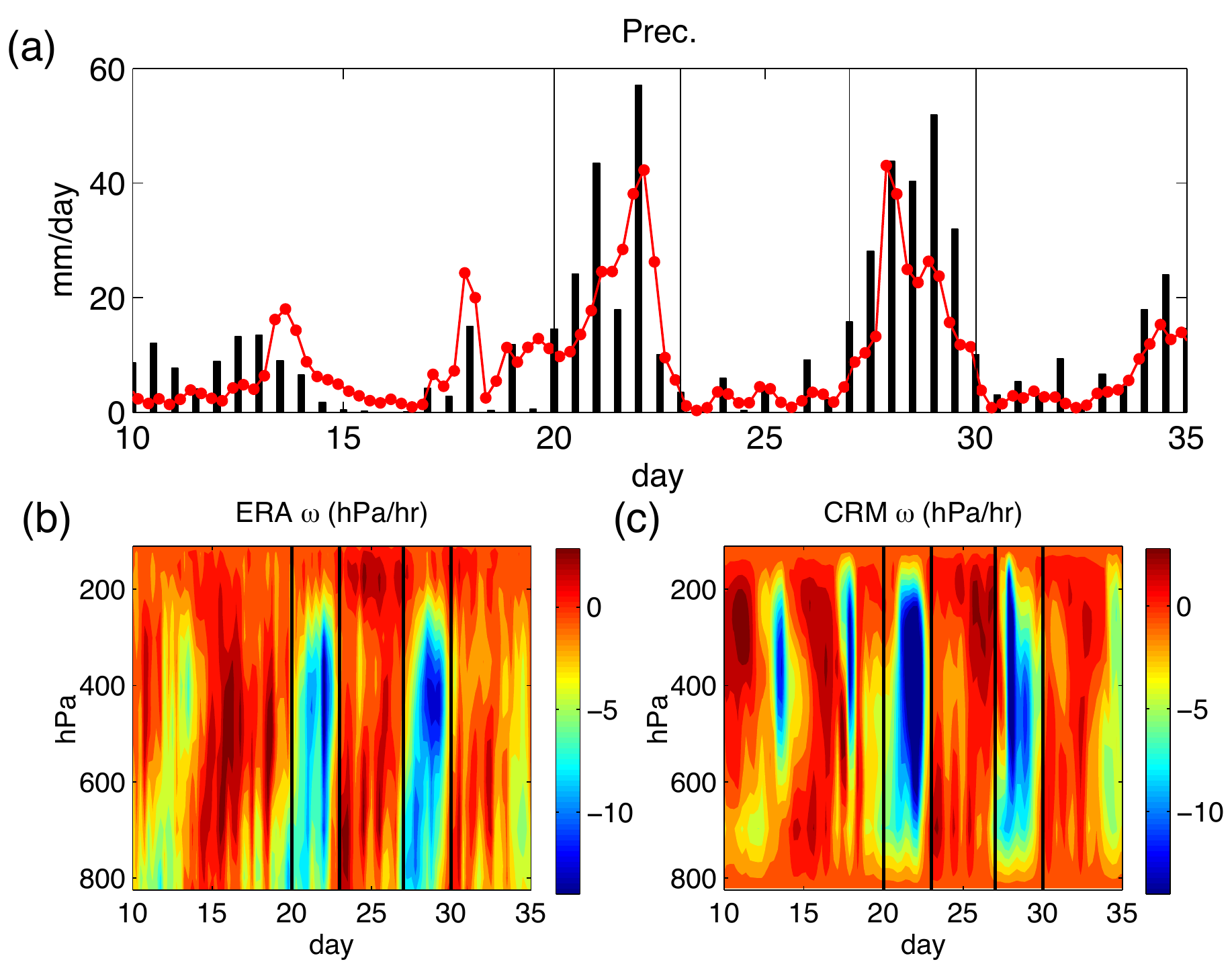} 
   \caption{(a) 12-hourly precipitation from the ERA-Interim reanalysis averaged over Northern Pakistan regional box (black bars), and CRM simulated precipitation in the control case (red dotted curve). (b) and (c) show the time series of $\omega$ in the ERA data and CRM control case as functions of time and pressure, respectively. Day 0 indicates July 1, 2010. The black vertical  lines mark day 20, 23, 27, and 30, indicating the time windows of the two extreme precipitation events.}
\end{figure}

\begin{figure}[htbp] 
   \centering
   \includegraphics[width=1 \textwidth]{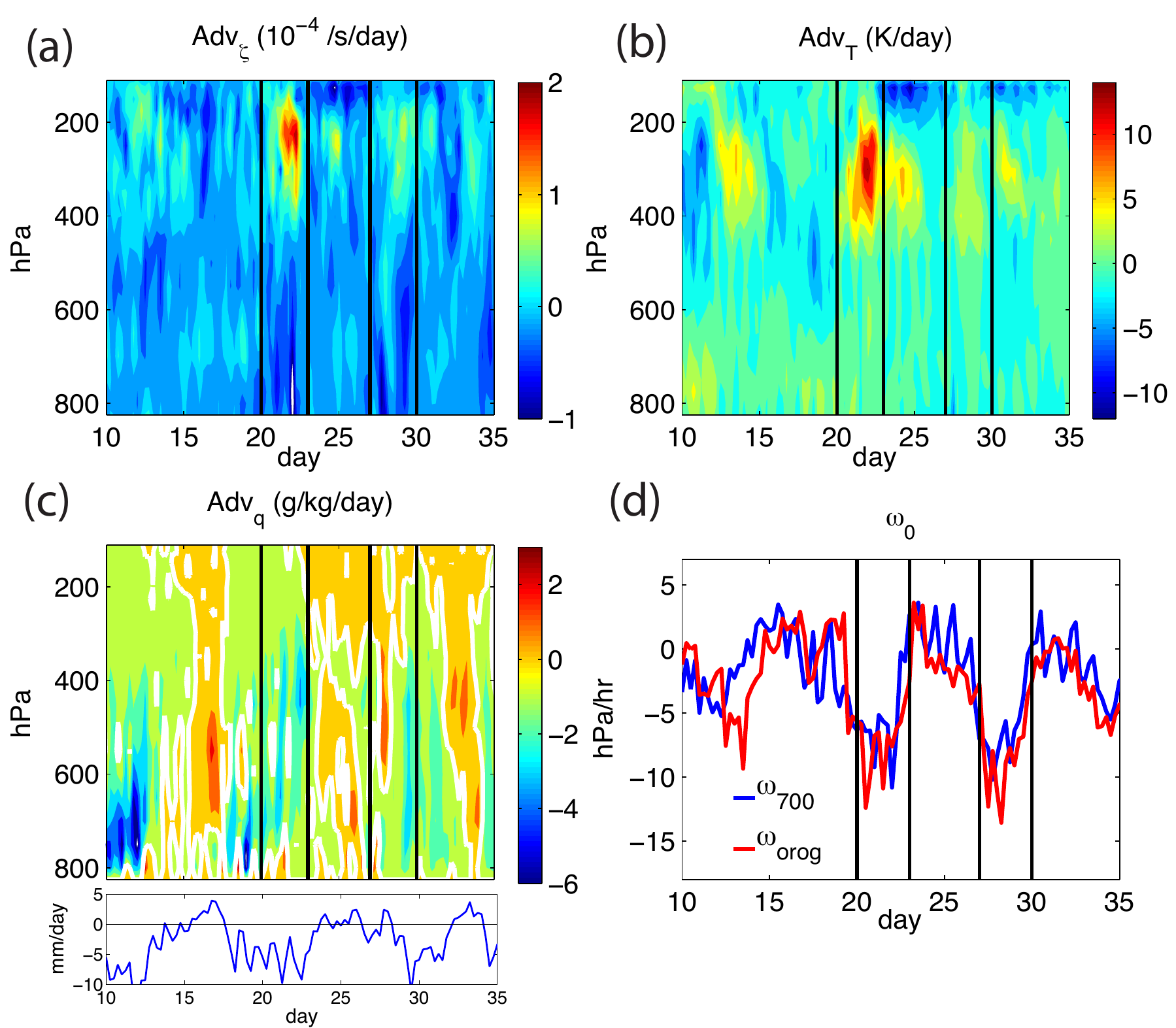} 
   \caption{Horizontal advection of (a) total vorticity, (b) temperature, and (c) moisture in the reanalysis data. For better visibility, the white contours in (c) are the zero contour lines. The lower panel in (c) plots the column-integrated horizontal moisture advection, in the unit of $mm\ day^{-1}$. (d) shows $\omega$ at 700 $hPa$ and the estimated orographic lift ($\omega_{orog}=\vec{V}_{0} \cdot  \nabla h_s$.)  }
\end{figure}

\begin{figure}[htbp] 
   \centering
   \includegraphics[width=1 \textwidth]{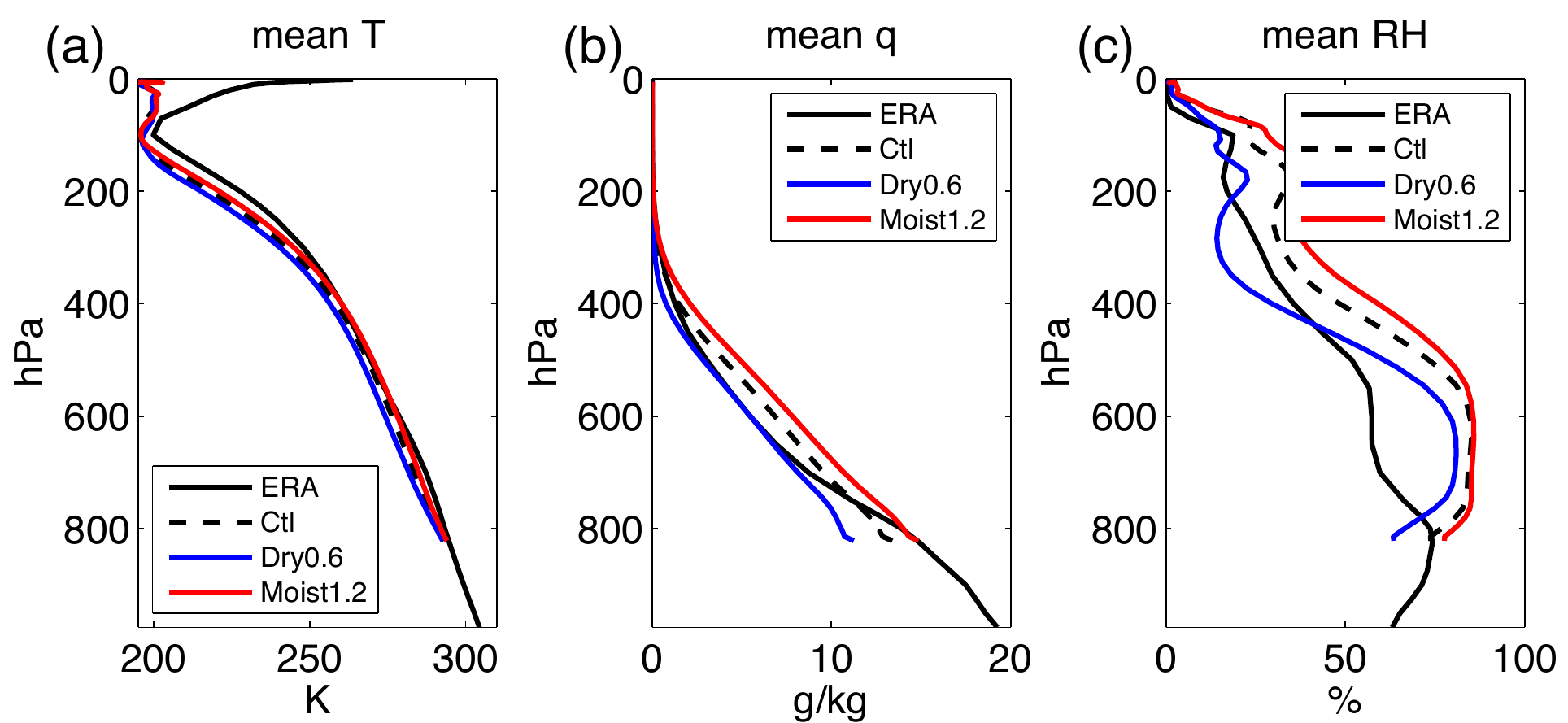} 
   \caption{The (a) temperature, (b) moisture, and (c) relative humidity profiles of reanalysis data and CRM results averaged between day 10 and day 40. The description of cases Dry0.6 and Moist1.2 is in section 3c. The profiles in the reanalysis data extend to 1000 $hPa$, however, only levels above the surface are included in the averaging.}
\end{figure}

\begin{figure}[htbp] 
   \centering
   \includegraphics[width=1 \textwidth]{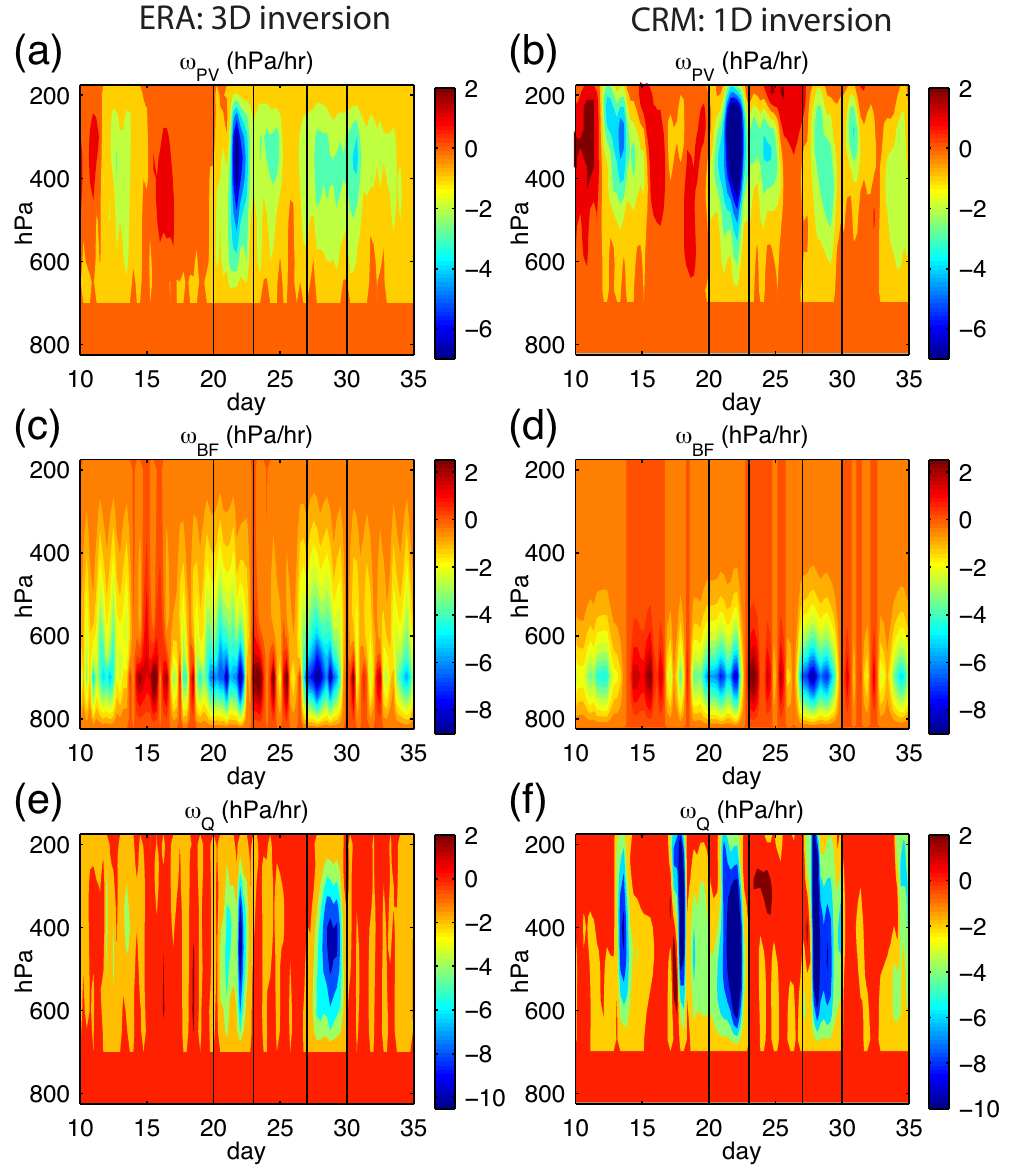} 
   \caption{ The decomposition of QG $\omega$. From top to bottom, plotted are $\omega$ due to PV forcing ($\omega_{PV}$), due to orographic lift ($\omega_{BF}$), and due to diabatic heating ($\omega_{Q}$). Note the color bars on different rows are different. The left column shows results from reanalysis data with three-dimensional QG$\omega$ inversion. The right column shows the results of the CRM control case with one-dimensional QG$\omega$ inversion.   }
\end{figure}

\begin{figure}[htbp] 
   \centering
   \includegraphics[width=1 \textwidth]{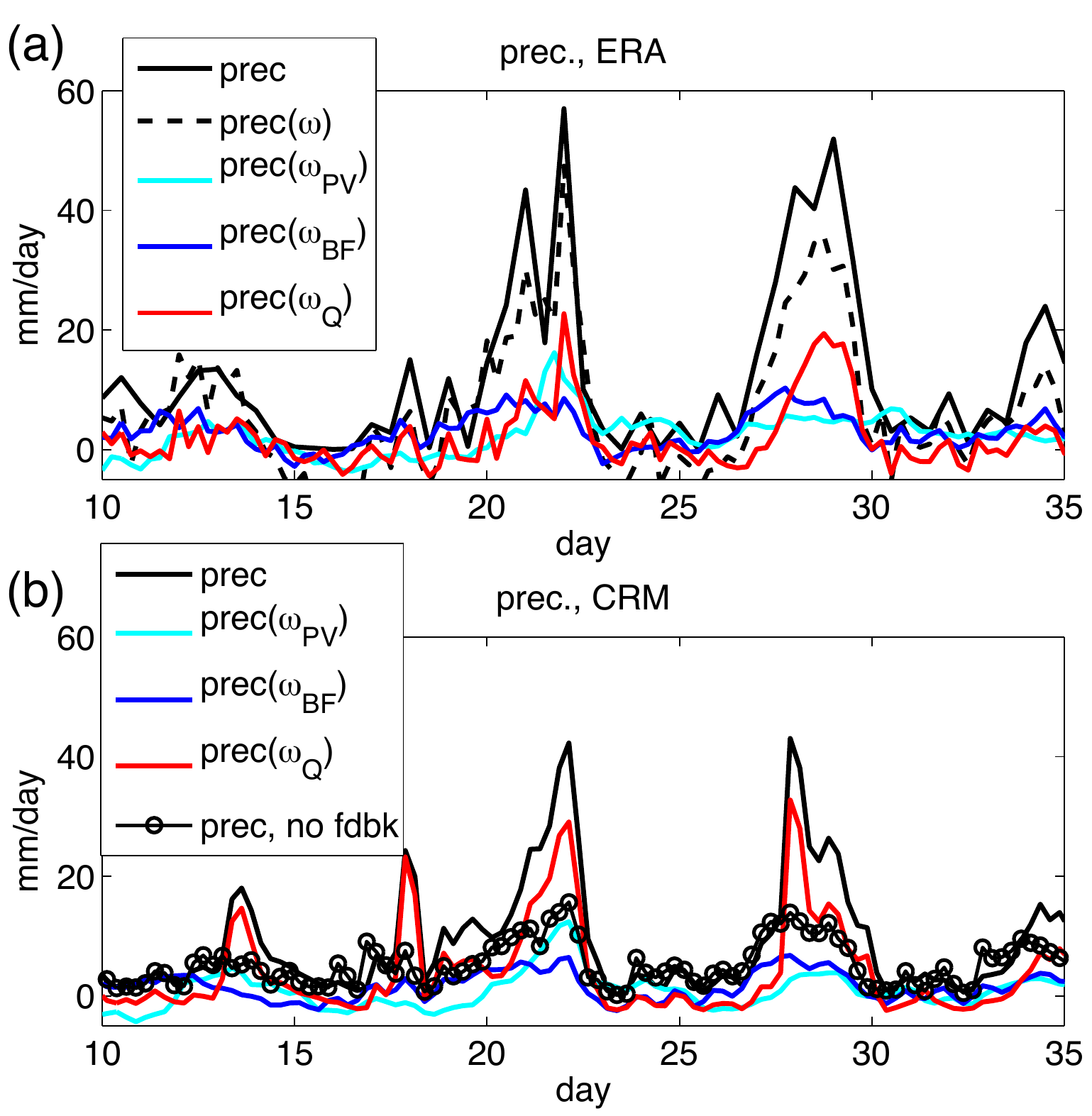} 
   \caption{The decomposition of precipitation on each $\omega$ component from (a) reanalysis data, and (b) the CRM control case. In (a), the black line is precipitation from reanalysis, and the black dashed line is precipitation converted from $\omega$, which is further decomposed into components    
associated with $\omega_{PV}$ (cyan line), associated with $\omega_{BF}$ (blue line); associated with $\omega_{Q}$ (red line). In (b), the black dashed line (sum of color lines) almost overlaps with the black solid line, thus is omitted. The black circle line in (b) is the CRM results with the diabatic heating feedback in the QG$\omega$ equation turned off.   }
\end{figure}

\begin{figure}[htbp] 
   \centering
   \includegraphics[width=1 \textwidth]{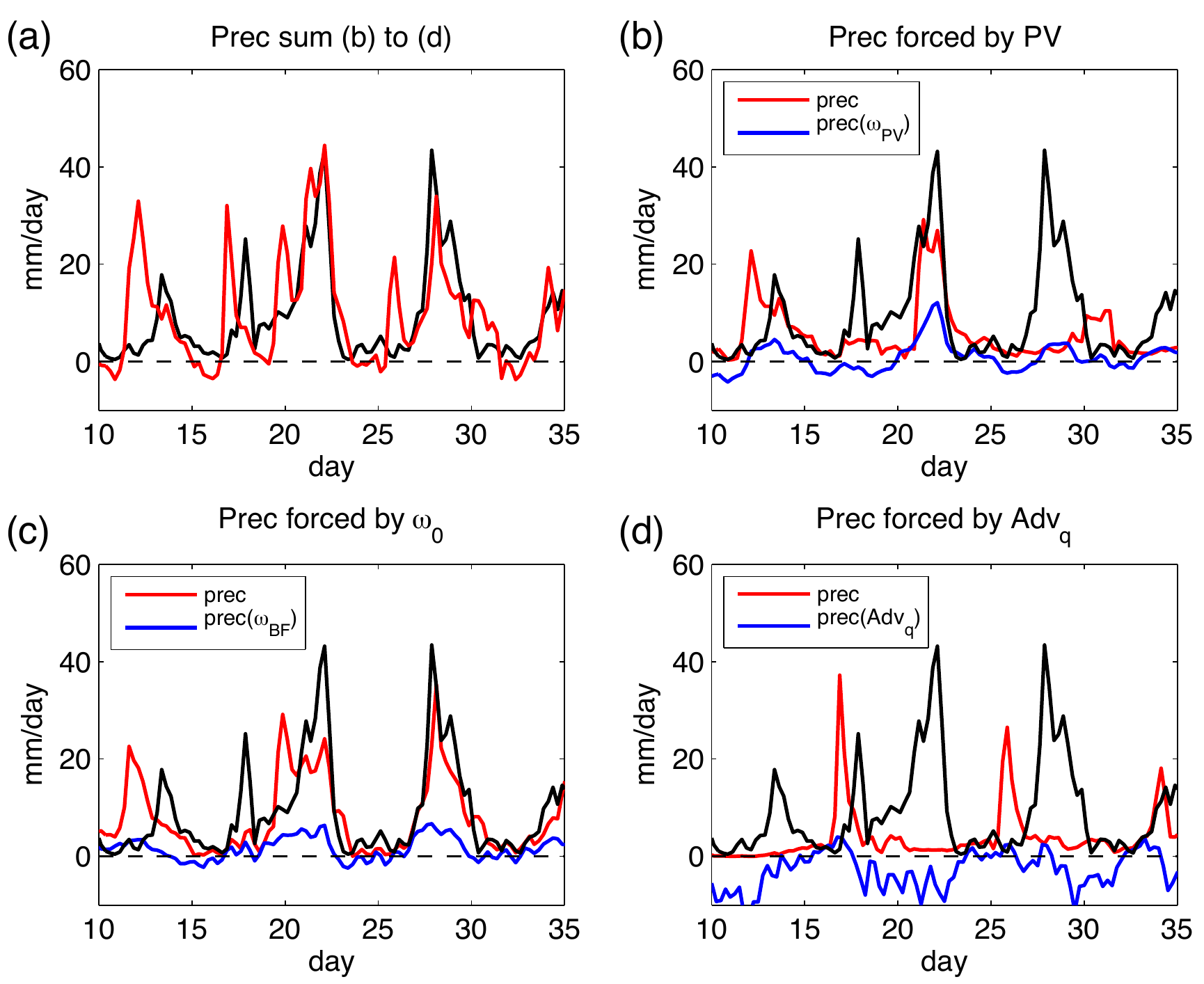} 
   \caption{(b)-(d): The red lines are CRM precipitation forced by only (b) PV forcing, (c) orographic lift, and (d) horizontal moisture advection. The blue lines are precipitation time series corresponding to each individual imposed forcing. The red line in (a) is the sum of the red lines in (b)-(d) with a subtraction of 8 $mm\ day^{-1}$ as multiple-counted mean precipitation. As references, the control case precipitation (black line) is also plotted in each panel. }
\end{figure}

\begin{figure}[htbp] 
   \centering
   \includegraphics[width=0.5 \textwidth]{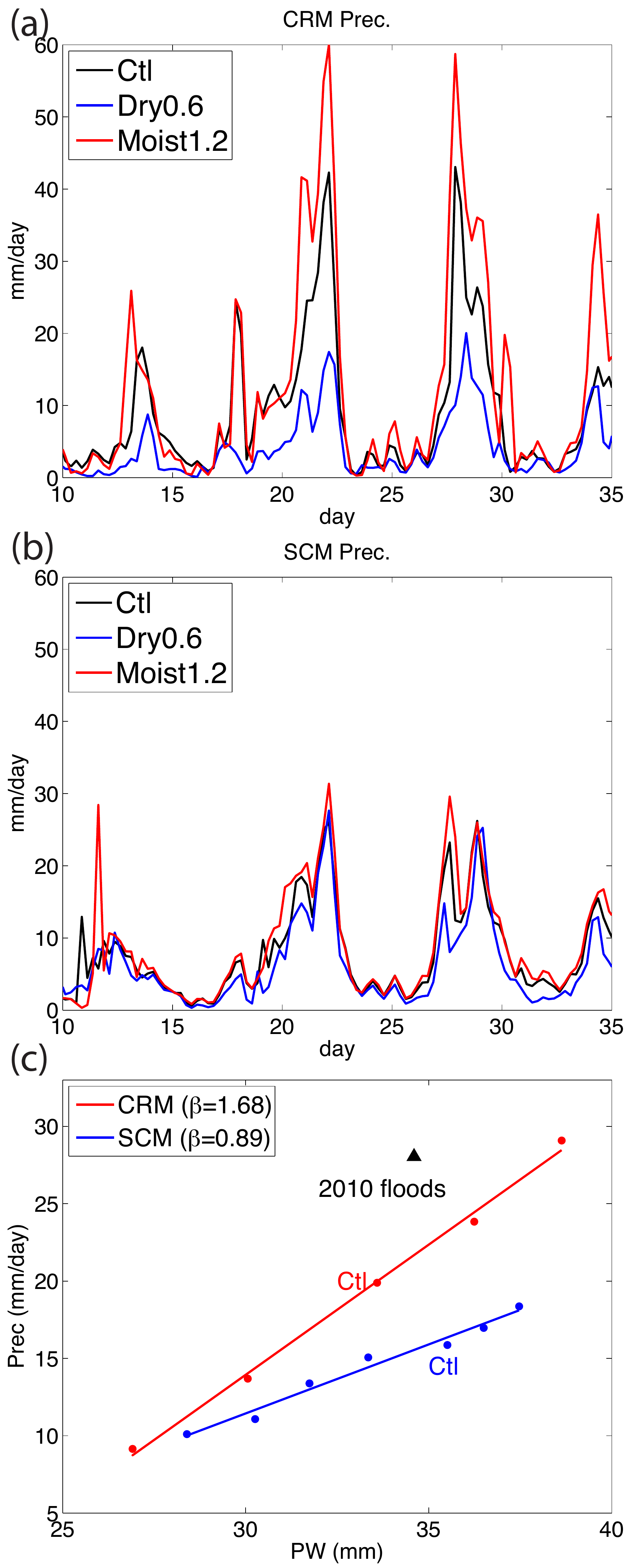} 
   \caption{The (a) CRM and (b) SCM simulated precipitation with different background relative humidity. (c) is a scatter plot of the precipitation during the events against the background precipitable water from all the cases. The color lines are the linear fit with slope shown in the legend. The black black triangle corresponds to the 2010 events from the reanalysis data. }
\end{figure}

\begin{figure}[htbp] 
   \centering
   \includegraphics[width=1 \textwidth]{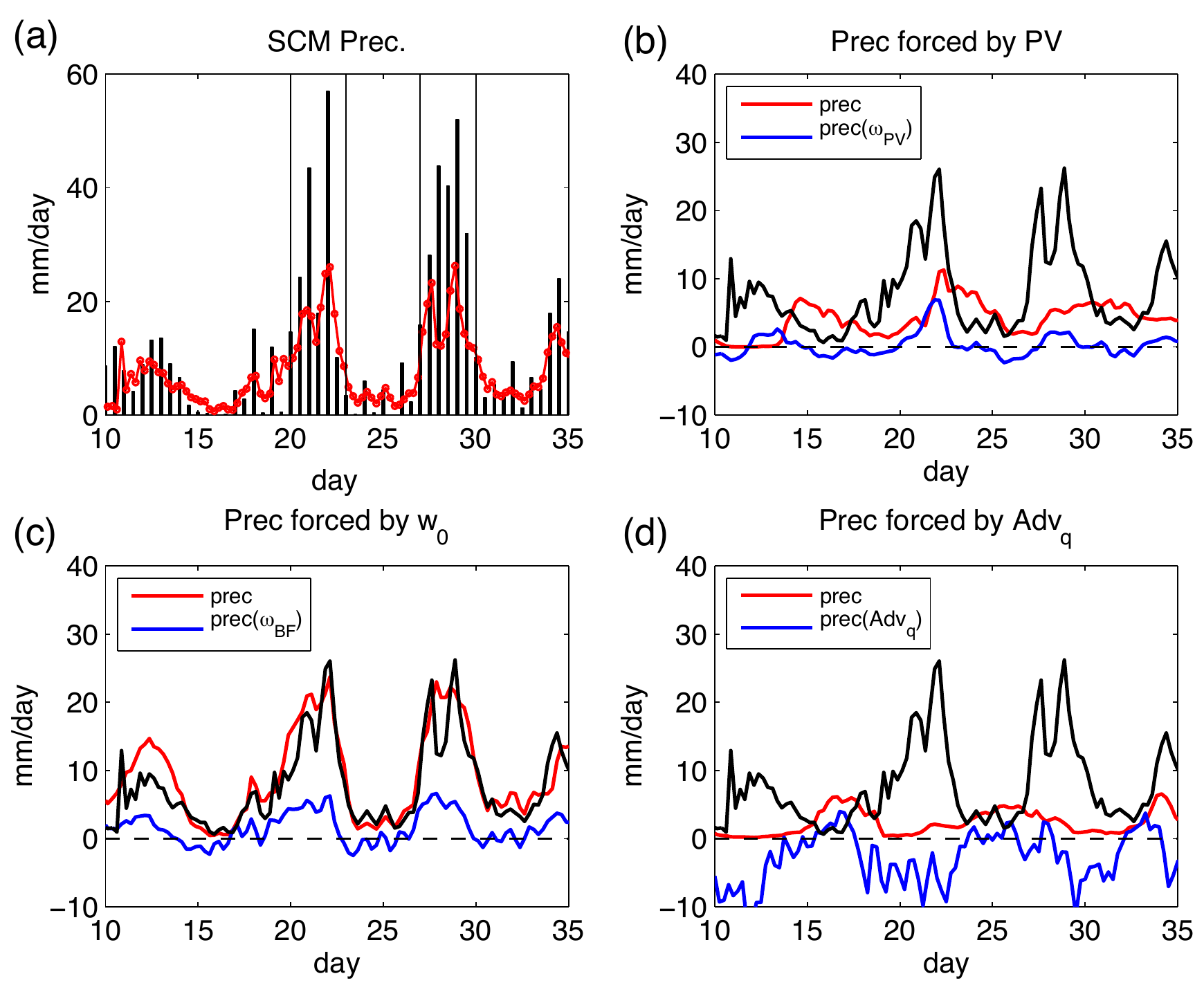} 
   \caption{ The SCM results. (a) is same with Fig. 1a, except the red dots showing the SCM results; (b)-(d) are same with Fig. 6b-d, respectively, except using the SCM. Note the y-axis in (b)-(d) are stretched for better visibility.   }
\end{figure}

\begin{figure}[htbp] 
   \centering
   \includegraphics[width=1 \textwidth]{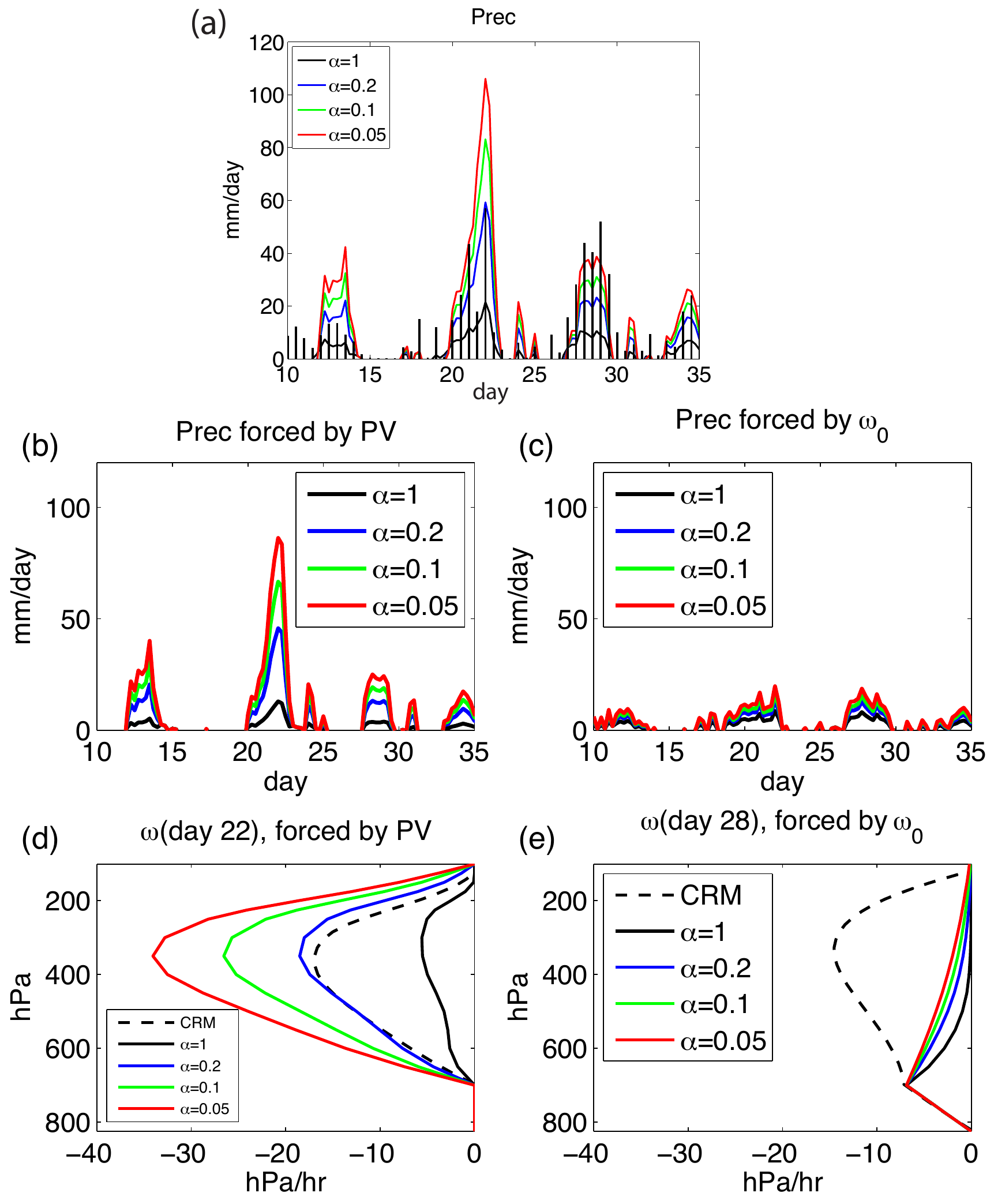} 
   \caption{Results with the dry QG$\omega$ equation with the use of $\sigma_e$ (Eq. 7). (a) is same with Fig. 1a, except the color lines showing results with different $\alpha$. The middle row shows precipitation forced by only (b) PV forcing and (c) orographic lift. (d) shows vertical profiles of $\omega$ forced by the PV forcing at day 22. (e) shows vertical profiles of $\omega$ forced by the orographic lift at day 28. In (d)-(e), the black dashed line shows the corresponding $\omega$ from the CRM simulations. }
\end{figure}

\end{document}